\documentstyle[aps,prl,multicol]{revtex} 
\begin{document}
\title{Non-equilibrium thermodynamic description of junctions in 
semiconductor devices}

\author{G. Gomila and J.M. Rub\'{\i}
\\Departament de F\'{\i}sica Fonamental\\Facultat de F\'{\i}sica\\
Universitat de Barcelona\\Diagonal 647, 08028 Barcelona, Spain}

\maketitle
\parskip 2ex

\begin{abstract}

The methods of non-equilibrium thermodynamics of systems with
an interface have been applied to the study of transport processes in 
semiconductor junctions.
A complete phenomenological model for drift-diffusion processes in a 
junction has been derived, which includes, from first principles, both 
surface equations and boundary conditions, together with the usual
 drift-diffusion equations for the bulks. In this way a 
self-consistent 
 characterisation
of the whole system, bulks and interface, has been obtained in a 
common
framework. The completeness of the model has been shown and a simple
application to metal-semiconductor junctions developed.

\end{abstract}
\begin{multicols}{2}
\section{Introduction}
The phenomenological approach to semiconductor device modelling has 
proven to 
be very powerful and successful during the last decades. By means of a
relatively
simple set of equations (i.e., balance equations,constitutive 
relations,
Maxwell equations), numerous devices have been modelled~\cite{Sze}. In
spite of
this success some aspects of the phenomenological models still remain 
not so well developed, namely the boundary conditions necessary to 
solve the set
of phenomenological equations and the appropriate equations to 
describe
the surface processes phenomenologically. These are, by no means, 
secondary
problems in semiconductor devices, since the behaviour of most of the 
devices
are critically affected by them. 
For instance, any device including a junction 
needs the appropriate boundary conditions, or discontinuity relations,
to be imposed at the junction, as well as, the
proper modelling of the superficial phenomena. In 
the term junction we also include the contacts used on the devices,
whose effects may sometimes be relevant, for instance, in the 
development
of semiconductor instabilities~\cite{Kromer}. 

Usually, these
problems are discussed more or less intuitively, by invoking sometimes
unclear physical arguments. We may  say that, at this moment,
in the current literature, a general
phenomenological theory for boundary conditions and surface effects 
is lacking, although some valuable work has been 
published in this area~\cite{Ancona}.
It seems, therefore, desirable to propose phenomenological models that
incorporate
both boundary conditions and surface equations in a fundamental and 
general
way. With such models, a self-consistent characterisation of 
interfaces
in semiconductor systems would then be able to be reached.
Therefore, for a given 
junction,
the models would provide the corresponding surface equations and 
boundary
conditions appropriate for it simply by adjusting a few surface 
parameters.
Clearly, from this point of view, semiconductor device modelling turns
out
 to be a much easier task. Moreover,
 only with this kind of approach, the presently accepted models
will be ultimately justified and, if necessary, improved in a 
systematic
and general way. 

As shown in~\cite{Ancona}, these kinds of models may be constructed
by means of non-equilibrium thermodynamic arguments. In this paper
we will also make use of such an approach, but following a rather 
different
formulation; namely, the one proposed in ~\cite{Mazur},
 in the context of fluid systems (see review~\cite{Bedo2} and 
references
   quoted therein). Non-equilibrium thermodynamic
derivation of transport models have been widely used in the past
for bulk systems~\cite{Degro}, usually in the context of fluid 
systems, but
 also in the context of semiconductor 
systems~\cite{Negre}. For systems with interfaces, much less work
has been done, although a rather complete theoretical treatment of 
this
problem has been already published~\cite{Bedo2}. 

In this paper we will 
present an application of such a formalism,
to the derivation of a transport model for drift-diffusion processes 
in a
junction.
As required above,
this model will incorporate, apart from the usual drift-diffusion 
equations for 
the bulk
systems forming the junction, the appropriate boundary conditions
and surface equations to respectively describe the coupling between 
the bulk 
systems and the surface processes.  As mentioned,
the entire model will be
obtained in a systematic way, starting from very general 
considerations about 
the non-equilibrium thermodynamics behaviour of the system. For this
reason, its generalisation
to include other irreversible processes may be obtained following the
same procedure as the one presented here. 

This paper is distributed as follows: in Section II we will discuss in 
detail the macroscopic description of the junction and the 
semiconductor
model we will use throughout the paper. In Section III we will present 
and 
derive
the phenomenological model, and consider some particularly interesting
cases.
The completeness of the model will be proven in Section IV, together
with a simple application to metal-semiconductor junctions. 
 Finally, in Section V we will summarize our main results.

\section{Macroscopic description of a junction}
From a macroscopic point of view, a junction may be defined as the 
very narrow 
region that
exists between two semiconductors in contact, and which exhibits 
peculiar 
macroscopic
behaviour reflected in the fact that the 
equations of state, transport coefficients, etc. are different from 
those of
the bulks. It is clear that due to the possibility of jumps in the 
relevant 
quantities, a description of the system in terms of variables varying 
continuously from one region to the other is not appropriate. 
Therefore,
 it is  
necessary to find the  jump relationships coupling the different 
parts, in terms
of the interfacial properties.

To this end we will introduce in our system a geometrical surface 
inside the 
 interfacial region that divides the system in two separate parts, 
referred to 
 as
 the bulk parts, which will be denoted by $-$ and $+$. We assume that 
each of
 these bulk parts behaves macroscopically as in the absence of
 interface. Of course this behaviour will be different from the {\em 
real}
 one inside the interfacial region. To take this fact into account one
 introduces {\em surface} variables, denoted by the superscript s, 
defined only 
 at the geometrical surface. These variables are the appropriate to 
describe
 the interfacial region.  

 The precise definition of these surface variables runs as follows.
 Let $\theta^{+}$ ($\theta^{-}$) be  Heaviside's functions whose 
values are 1 (0) 
 for points located on the right (left) of the surface and 0 (1) for 
the 
 opposite case. Let $d$ be an unspecified field and $d^{\pm}$ be the 
 corresponding
 fields in the bulks, referred to as the {\em extrapoled} fields. As 
 pointed 
 out before, these fields differ from the real ones in 
 the interfacial region. To account for this difference we introduce 
 {\em excess} fields $d_{ex}$ as follows~\cite{Bedo2} 
 \begin{equation}
 d_{ex} (\vec{r}) = d (\vec{r}) - d^{+}(\vec{r}) \theta^{+}(\vec{r}) 
                                - d^{-}(\vec{r}) \theta^{-}(\vec{r})
  \label{dexces}                              
 \end{equation}
 These excess fields have, by definition, two important properties: 1)
  They vanish away from the interfacial region, because there
  the effects of the interface die out, and consequently the {\em 
real} 
  fields and the {\em bulk} ones 
 coincide. Thus the 
 excess fields are only defined in the interfacial region. 2) They 
take
 small values when
 interfacial effects are small, that is, when the {\em real} fields do
 not depart significantly from the {\em extrapoled} ones.
  
 The {\em surface } variables are defined as averages of the excess 
fields. 
 For a planar surface at rest identified with the plane $x =0$, we can 
write
 ~\cite{Bedo4}
 \begin{equation}
 d^{s}(y,z) = \int dx \, d_{ex}(\vec{r})
 \label{surface}
 \end{equation}
 where the integration may be extended outside of the interfacial 
region,
  for 
 the excess fields vanish very quickly there.
 These fields constitute the relevant quantities in the description of 
 continuum systems in the presence of interfaces, as was first pointed
 out in~\cite{Gibbs}. 
 
 Note that, with this characterisation, the system is seen as composed
by three
 sub-systems: two bulks, with unchanged properties just until the 
geometrical
 interface and  described by means of the extrapoled variables,
 and an interface, accounting for the properties
 of the interfacial region through the surface variables.
 
 As noted before, in this paper we will follow a thermodynamic 
approach. This
 forces us to give an a priori thermodynamic characterisation of
 the systems involved. 
 Here we will assume that the bulk sub-systems
  belong to the class
 of the multicomponent electrically conducting polarizable fluids
  whose thermodynamic variables are: the densities of internal energy,
$u$, 
  and 
  entropy, $s$, the number density of each component, $n_{k}$, 
  and the electric polarisation $\vec{P}$
  (magnetic polarisation, $\vec{M}$, may also be included). 
  The surface sub-system will belong to the same class, but neglecting
  any polarisation phenomena.
  The bulk system densities will be defined per unit of volume while 
at
  the surface per unit of area. 
   
 This  semiconductor fluid-like model we will use throughout this 
 paper, is very related 
 to the one proposed 
 in~\cite{Ancona} in which a semiconductor may be considered as the 
 superposition of three continuum media, namely the crystal lattice, 
the 
 electrons
  and the holes. The first one is, in general, positively or 
negatively charged
  depending on the sign of the ionised impurities. It will be 
considered
  at rest with fixed mass and impurity densities, not including for 
example 
  any aging 
  phenomena. This sub-system is mainly responsible for the 
polarisation 
  effects. Electrons and holes will be considered charged fluids. For 
the sake
  of simplicity, we will assume that each component has the same
  charge and mass on each side of the junction. This assumption
  considerably simplifies our analysis in the sense that we are 
reducing the
  number of constituents in our description.
  
  It is worth nothing that although the formalism we present applies 
to 
 semiconductor systems,
 it can also be used to study other kinds of materials, like metals or
 insulators, because from the
 thermodynamic point of view, they have similar characteristics. It is
only 
 when the state equations are specified that 
 the differences become substantial.

 \section{Non-equilibrium Thermodynamics model for a junction}
 
 Once the thermodynamic description of the junction has been 
established, we 
 can now proceed
 to  investigate
 the non-equilibrium behaviour of the whole system by applying the 
usual methods
 of non-equilibrium thermodynamics~\cite{Degro}. Our main purpose in 
this 
 section will be to 
 obtain the constitutive relationships for the 
 irreversible processes in the bulk and at the interface together with
boundary
 conditions coupling the different parts of the system. As noted 
before,
 a similar approach was previously
 proposed in~\cite{Bedo} in the context of fluid systems.
 
 Although the calculations may be worked out in the very general case,
by 
 including all possible irreversible processes as in~\cite{Bedo}, we 
will
 consider, for the sake of simplicity and clarity, only the drift 
 and diffusion processes, neglecting the
 recombination or generation processes, magnetic effects, thermal 
processes 
 or relaxation
 polarisation processes. The last three may be introduced in a more or
less
 straightforward way by following the procedure presented in this 
paper. 
 Generation-recombination processes deserve special treatment in the
 framework of non-equilibrium thermodynamics, due to the 
 intrinsic non-linear nature of the constitutive 
relations~\cite{Degro}.

 Moreover, we will also consider a one-dimensional system in 
 which 
 significant variations of the relevant quantities used in its 
description occur 
 only in a direction perpendicular to the interface. Vectorial fields 
will 
 then have components along the normal to the junction, and the 
vectorial 
 notation will be suppressed.

\subsection{Basic equations of the model} 

This subsection is devoted to the introduction of the complete set of 
equations that constitutes
the model, leaving its rigorous derivation for next subsection.
As mentioned in the introduction, the equations of the model may be 
divided 
in three 
categories making reference to bulk, surface, and boundary conditions, 
respectively:

\vspace{0.3 cm}

i) {\em Bulk equations}. The equations describing the bulks consist of
balance
 equations, constitutive
relations and Poisson's equation, supplemented by the corresponding 
equations of state. They have been found to be
\begin{eqnarray}      
\mbox{}& &\frac{\partial n_{k}^{\pm}}{\partial t} + 
      \frac{\partial J_{k}^{\pm}}{\partial x} = 0
     \hspace{1 cm} \mbox{;} \hspace{0.5 cm} k=1,2 \label{contk}\\
\mbox{}& &\frac{\partial n_{3}^{\pm}}{\partial t}= 0 \label{cont3}\\
\mbox{}& &\frac{\partial D^{\pm}}{\partial x} = q^{\pm} \label{qPo}\\
\mbox{}& & J_{k}^{\pm} = - 
D_{k}^{\pm} \frac{\partial n_{k}^{\pm}}{\partial x}+
         (-1)^{k} \tilde{\mu}_{k}^{\pm} n_{k}^{\pm} E^{\pm} 
\hspace{0.5 cm} 
         \mbox{;} \hspace{0.5 cm} k=1,2
          \label{driftd} 
\label{balm+1}
\end{eqnarray}
where $n_{k}^{\pm}$ is the particle number density with the index $k$ 
referring
to the 
subcontinua electrons ($k=1$), holes ($k=2$) and lattice ($k=3$), 
$q^{\pm} = \sum_{k=1}^{3} q_{k} n_{k}^{\pm}$ is the total density 
charge, with
$q_{k}$ the charge of the $k$-th component,  
 $D^{\pm}$ is the electric displacement, $J_{k}^{\pm}$ is the particle
number
 density
 current, $D_{k}^{\pm}$ is the
 diffusion coefficient of $k$-th component, and 
$\tilde{\mu}_{k}^{\pm}$ 
 the corresponding 
 mobility. These two last quantities are related through 
 Einstein's relation
 \begin{equation}
 \frac{D_{k}^{\pm}}{\tilde{\mu}_{k}^{\pm}} = \frac{\partial 
 \mu_{k}^{\pm}}{\partial n_{k}^{\pm}}\frac{n_{k}^{\pm}}{e}   
 \hspace{0.5 cm} \mbox{;} \hspace{0.5 cm} k=1,2
 \label{Einstein}
 \end{equation}
 where $e$ is the (positive) electron charge.
 These bulk equations constitute the usual drift-diffusion model for a
 semiconductor when generation-recombination processes are 
neglected~\cite{Sze}
  (Eq.~(\ref{cont3}) simply states that the aging phenomena are 
neglected).
 Eq.~(\ref{driftd}) can be rewritten in a more appealing form by 
introducing 
 the electrochemical 
 potential for the $k$th component as
 \begin{equation}
 \phi_{k}^{\pm} = \mu_{k}^{\pm} + q_{k} V^{\pm} \hspace{0.5 cm} 
\mbox{;}
  \hspace{0.5 cm}  k = 1 , 2 
 \label{elecpot}
  \end{equation} 
 with the electric potential defined as usual by $E^{\pm} = - 
 \partial V^{\pm} /\partial x$. For the constitutive relations, one 
then
 obtains
 \begin{equation}
 e J_{k}^{\pm} = - n_{k}^{\pm} \tilde{\mu}^{\pm}_{k} 
                \frac{\partial \phi_{k}^{\pm}}{\partial x}
      \hspace{0.4 cm} \mbox{;} \hspace{0.4 cm} k=1,2
\label{driftd1}
\end{equation}
which shows that, for drift-diffusion processes, the electrochemical 
potentials 
(usually called quasi-Fermi levels)
play an important role,  as we will
see later when boundary conditions are expressed in a natural way in 
terms 
of these potentials.

The appropriate equation of 
state for the bulk systems corresponding to the model presented are 
\begin{eqnarray}
\mu_{k}^{\pm} & = & \mu_{k}^{\pm} (n_{k}^{\pm}) 
\hspace{0.5 cm} \mbox{;} \hspace{0.5 cm} k=1,2 \label{stat1}\\
       D^{\pm}& = &\epsilon^{\pm} E^{\pm} \label{stat2}
\end{eqnarray}
where $\epsilon^{\pm}$ are the bulk dielectric permitivities, assumed 
to be independent of $n_{k}^{\pm}$. For the case of 
non-degenerate semiconductors one has for the first state equations
\begin{equation}
\mu_{k}^{\pm} =  \mu_{k,0}^{\pm} + 
   k_{B}T \ln \left( \frac{n_{k}^{\pm}}{N_{k}^{\pm}} \right) 
   \hspace{0.5 cm} \mbox{;} \hspace{0.5 cm} k=1,2 \label{nonde}
\end{equation}
where $\mu_{k,0}^{\pm}$ is a chemical potential of reference, with 
$N_{k}^{\pm}$ being
the effective density of states in the conduction ($k=1$) and valence 
($k=2$)
band. In that case Einstein's relation 
reduces to its usual expression~\cite{Sze},
\begin{equation}
\frac{D_{k}^{\pm}}{\tilde{\mu}_{k}^{\pm}} = \frac{k_{B} T^{\pm}}{e}
\hspace{0.5 cm} \mbox{;} \hspace{0.5 cm} k=1,2  \, .
\end{equation}

\vspace{0.5 cm}

ii) {\em Surface equations}. As for the bulk, surface equations would 
consist 
in general of 
balance equations, constitutive relations and state equations. For the
case 
of a one-dimensional
system no surface constitutive relations are needed, because no 
transport 
process takes place along the surface. 
In that case, we then simply have,
 \begin{eqnarray}      
& &\frac{\partial n_{k}^{s}}{\partial t} +
      \left[ J_{k} \right]_{-} = 0
         \hspace{0.7 cm} \mbox{;} \hspace{0.6 cm} k=1,2  
\label{mas1dk} \\
& & \frac{\partial n_{3}^{s}}{\partial t}  = 0 \label{mas1d3} 
\end{eqnarray}
where $n_{k}^{s}$ is the surface density number of the kth component,
 and where we have
defined $[A]_{-} \equiv A^{+}-A^{-}$, with $A^{\pm}$ being evaluated
at the interface. Again Eq.~(\ref{mas1d3}) states that no aging 
phenomena 
are allowed, now at the interface.
Moreover, the surface state equations are in that case simply
\begin{equation}
\mu_{k}^{s} = \mu_{k}^{s}(n_{k}^{s})  
\hspace{0.5 cm} \mbox{;} \hspace{0.5 cm} k=1,2 \, .\label{stat1s}
\end{equation}
Remember that we are assuming that the interface is non-polarizable 
and hence
no surface state equation for the surface polarisation is necessary 
(see
appendix) 

\vspace{0.3 cm}

iii) {\em Boundary conditions}. For the boundary conditions, we have 
to distinguish between
the ones corresponding to Poisson's equation and the ones that 
correspond to the
drift-diffusion processes. The former ones may be written as
\begin{eqnarray}
\mbox{} [D]_{-} & = & q^{s} \label{discP} \\
\mbox{} [V]_{-} & = & 0 \label{En}
\end{eqnarray}
where $q^{s} = \sum_{k=1}^{3} q_{k} n_{k}^{s}$ is the net surface 
charge,
whereas for the latter we have obtained 
\begin{equation}
J_{k}^{\pm} = 
           -\frac{1}{T} F_{k}^{\pm} ( \{ \phi_{j}^{+}-\phi_{j}^{s}\},
                     \{\phi_{j}^{s}-\phi_{j}^{-}\} ) \hspace{0.3 cm} 
                     \mbox{;} \hspace{0.2 cm} k,j=1,2 
\label{F+-}
\end{equation}
where $\phi_{k}^{s} = \mu_{k}^{s} + q_{k} [V]_{+}$ is the 
electrochemical 
potential of the $k$-th component at the surface, with $[A]_{+} =
1/2 (A^{+} + A^{+})$, and where all the 
 bulk magnitudes are evaluated at the interface, that is, at $x=0$.

These equations can be 
formulated in a more compact way, by introducing a simplified notation
for the
differences of electrochemical potentials, $X_{k}^{+} \equiv 
\phi_{k}^{+}-\phi_{k}^{s}$,  
 $X_{k}^{-} \equiv \phi_{k}^{s}-\phi_{k}^{-}$, called, from now on,
surface thermodynamic forces, and for the fields corresponding 
 to the same quantity but for 
different constituents, for instance, 
$\vec{J}^{\pm} = ( J_{1}^{\pm}, J_{2}^{\pm} )$ where 1 and 2 refer to
electrons and holes respectively. The scalar product of two 
such vectors will then read
$\vec{J}^{+} \cdot \vec{X}^{+} =
\sum_{k=1}^{2} J_{k}^{+} X_{k}^{+}$. In this notation, Eq.~(\ref{F+-})
can be
 written simply as
\begin{equation}
 \vec{J}^{\pm} = -\frac{1}{T} \vec{F}^{\pm} ( \vec{X}^{+}, 
\vec{X}^{-}) \, .
 \label{F+}
\end{equation}
The boundary conditions for the drift-diffusion processes, 
Eq.~(\ref{F+-}) or
Eq.~(\ref{F+}), deserve some comments.
First of all, it should be noted that they are 
non-equilibrium boundary conditions, that is, they relate the 
discontinuities in the relevant quantities in the drift-diffusion 
process,
namely the electrochemical potentials to the currents. Moreover, 
although
they have been formulated in terms of currents and forces (see above),
 they may also depend on the
surface thermodynamic variables (see next subsection). Finally,
the functions $F_{k}^{\pm}$ are subject to 
certain thermodynamic restrictions, namely 
\begin{eqnarray}
& & \vec{F}^{\pm}( \vec{0}, \vec{0} )= 0  
 \label{equil} \\
& &\vec{X}^{+} \cdot \vec{F}^{+} + \vec{X}^{-} \cdot \vec{F}^{-}  \ge 
0 
\label{nonequil} 
\end{eqnarray}

\begin{equation}
\left. \frac{\partial F_{k}^{\alpha}}{\partial X_{j}^{\beta}} 
\right|_{eq} =
\left. \frac{\partial F_{j}^{\beta}}{\partial 
X_{k}^{\alpha}} \right|_{eq} \hspace{0.4 cm} {k, \, j = 1, 2 \atop 
\alpha, \, \beta = +, -} 
\label{Onsager}
\end{equation} 
coming from the equilibrium condition, the second principle
of thermodynamics, and the Onsager's symmetries,
respectively. Note that we have $n (n-1) /2 $ symmetry relations, 
where
n is the number surface thermodynamic forces (in the general case n= 
4). 

\vspace{0.3 cm}

It should be noted that for the model presented above,
the total electric density current 
is a continuous field through the interface. Indeed, from 
Eqs.~(\ref{mas1dk})
and~(\ref{mas1d3}), the total surface charge 
equation for a one-dimensional system is found to be
\begin{equation} 
\frac{\partial}{\partial t} q_{A}^{s} + [ J_{c} ]_{-} = 0
\label{baltqs1d}
\end{equation}
where $J_{c}^{\pm} = \sum_{k = 1}^{3} q_{k} J_{k}^{\pm}$, is the total
electric conducting 
density current.
Upon substituting Eq.~(\ref{discP}) in Eq.~(\ref{baltqs1d}), we obtain
\begin{equation}
[I]_{-} = 0
\label{Icon}
\end{equation}
where 
\begin{equation}
I^{\pm} = \frac{\partial}{\partial t} D^{\pm} + J_{c}^{\pm}
\label{Amper}
\end{equation}
is the total electric density current which can be shown, from
the semiconductor bulk equations, to only be a function of time
(Ampere's law). This fact, and Eq.~(\ref{Icon}), makes the use 
of a superindex for this quantity irrelevant.

\subsection{Entropy production and constitutive equations}  

Our purpose in this subsection is to justify the equations proposed in
the previous subsection.

i) For the bulk equations we only have to derive the constitutive 
relations,
 Eq.~(\ref{balm+1}), and
 Einstein's relation Eq.~(\ref{Einstein}), because
Eqs.~(\ref{contk}) and~(\ref{cont3}) are simply continuity equations 
and 
Eqs.~(\ref{stat1}), (\ref{stat2}), appropriate state equations. More 
general
 equations of state would be possible a priori, but we will conclude 
from
 our analysis that these are the consistent ones with the remaining 
equations 
 of the model.
To derive the constitutive relations, we will analyse the bulk entropy
productions using the methods of non-equilibrium 
thermodynamics~\cite{Degro}.
In the case concerning 
us, that is when only drift-diffusion processes are considered,
 the entropy production reduces to (see appendix)
\begin{equation}
\sigma_{s}^{\pm} = - \frac{1}{T} \sum_{k=1}^{3} J_{k}^{\pm '} 
 \left( \frac{\partial \mu_{k}^{\pm}}{\partial x} - q_{k} E^{\pm} 
\right)
\label{elect+}
\end{equation}
where the quantities $J_{k}^{\pm '}$ are diffusion currents defined 
in the appendix.
After applying Prigogine's theorem~\cite{Degro}, for which we can
 redefine
the diffusive fluxes with respect to any desired velocity (in our case
we will
take the reference frame fixed to the lattice continuum which is 
assumed
to be at rest), Eq.~(\ref{elect+}) transforms into
\begin{equation}
\sigma_{s}^{\pm} = - \frac{1}{T} \sum_{k=1}^{2} J_{k}^{\pm}  
    \left( \frac{\partial \mu_{k}^{\pm}}{\partial x} - q_{k} E^{\pm} 
\right) \, .
\end{equation}
From it we can infer the  phenomenological constitutive relations. For
 bulk processes they are well represented through linear relations, 
that is,
\begin{equation}
J_{k}^{\pm} = -\frac{1}{T} \sum_{j=1}^{2} 
 L_{kj}^{\pm} \left(\frac{\partial \mu_{k}^{\pm}}{\partial x}- 
                                                         q_{k} E^{\pm}
\right)
      \hspace{0.3 cm} \mbox{,} \hspace{0.2 cm} k=1,2
 \label{Jbulk}
\end{equation}
where $L_{ij}^{\pm}$ is the matrix of phenomenological coefficients 
which
according to Onsager's principle is symmetric and according to second
law positive definite~\cite{Degro}, and only depends
  on the
bulk thermodynamic variables, which in our case are, for instance, 
$T^{\pm},
n_{k}^{\pm},E^{\pm}$. To arrive at the form given in 
Eq.~(\ref{driftd}) we will 
suppose that the cross effects
are negligibly small,  $L_{12}^{\pm}=L_{21}^{\pm} \approx 0$ and that 
the
 phenomenological 
coefficients and the chemical potentials do not depend on $E^{\pm}$. 
We will 
also
assume that the chemical potentials depend only on the corresponding 
density,
that is $\mu_{k}^{\pm} = \mu_{k}^{\pm} ( n_{k}^{\pm})$. Then defining 
the 
diffusion coefficient and the mobility for the electrons and holes as
\begin{eqnarray}
           D_{k}^{\pm} & = & \frac{L_{kk}^{\pm}}{T} 
        \frac{\partial \mu_{k}^{\pm}}{\partial n_{k}^{\pm}} 
\hspace{0.6 cm} 
        \mbox{;} \hspace{0.4 cm} k=1,2 \\
 \tilde{\mu}_{k}^{\pm} & = & \frac{L_{kk}^{\pm}}{T} 
        \frac{e}{n_{k}^{\pm}}  \hspace{0.6 cm} 
        \mbox{;} \hspace{0.4 cm} k=1,2 \, ,
        \label{mob}
\end{eqnarray}
Eq.~(\ref{Jbulk}) transforms into the drift-diffusion relation given 
in Eq.~(\ref{driftd}).

Einstein's relation, Eq.~(\ref{Einstein}), follows directly from the 
previous 
definitions of the diffusion and  mobility coefficients.
On the other hand, taking into account the definition of the 
electrochemical 
potential given in Eq.~(\ref{elecpot}), we may rewrite 
Eq.~(\ref{Jbulk}) as
\begin{equation}
J_{k}^{\pm} = -\frac{1}{T} \sum_{j=1}^{2} 
      L_{kj}^{\pm} \frac{\partial \phi_{k}^{\pm}}{\partial x}
      \hspace{0.3 cm} \mbox{;} \hspace{0.3 cm} k=1,2
\end{equation}
which after neglecting cross effects and considering the definition of
the 
mobility given through Eq.~(\ref{mob}), transforms into 
Eq.~(\ref{driftd1}).

\vspace{0.5 cm}

ii) The surface equations, Eqs.~(\ref{mas1dk}) and~(\ref{mas1d3}) are 
the 
corresponding
one-dimensional surface continuity equations and they have been 
derived in the 
appendix . As before, Eq. (\ref{stat1}) corresponds to 
particular forms of the
surface state equations appropriate to the model.

\vspace{0.5 cm}

iii) Finally, let us focus on the boundary conditions.
Eqs.~(\ref{discP}) and~(\ref{En}) are the usual discontinuity relation
for 
Poisson's equation and they do not deserve further comments (see the
appendix for the case in which surface polarisation is included in the
formulation and~\cite{Bedo4} for a general treatment of Maxwell's 
equations). 

To derive the remaining boundary conditions, Eq.~(\ref{F+-}), we 
proceed, as we 
did in the bulk, by 
analysing the surface entropy production.
It is shown in the appendix that for the case of drift-diffusion 
processes,
it may be written as
\begin{equation}
\sigma_{s}^{s} = -\frac{1}{T} \sum_{k=1}^{2} [J_{k} (\mu_{k} - 
           \mu_{k}^{s})]_{-} \, .
\label{s1}           
\end{equation}
Using the definition of the bracket $[ \cdot  ]_{-}$ we may 
rewrite Eq.~(\ref{s1}) as
\begin{equation}
\sigma_{s}^{s} = -\frac{1}{T} \sum_{k=1}^{2} J_{k}^{+} (\mu_{k}^{+} - 
           \mu_{k}^{s})  
                 -\frac{1}{T} \sum_{k=1}^{2} J_{k}^{-} (\mu_{k}^{s} - 
           \mu_{k}^{-}) \, .
\label{metentro}
\end{equation}
Finally, by using Eq.~(\ref{En}) and the definition of the surface
and bulk electrochemical potentials given in the previous subsection,
we arrive at our final expression for the surface entropy 
production,
 \begin{eqnarray}
\sigma_{s}^{s} &=& -\frac{1}{T} \sum_{k=1}^{2} J_{k}^{+} (\phi_{k}^{+}
- 
           \phi_{k}^{s})  
             -\frac{1}{T} \sum_{k=1}^{2} J_{k}^{-} (\phi_{k}^{s} - 
           \phi_{k}^{-}) \\
               &=& -\frac{1}{T} \vec{J}^{+} \cdot \vec{X}^{+}  
             -\frac{1}{T} \vec{J}^{-} \cdot \vec{X}^{-} \, .
 \label{metentr1}
\end{eqnarray}
This last expression enables us to identify  
the surface currents, $\vec{J}^{\pm}$, and their conjugated
 surface thermodynamic 
forces, $\vec{X}^{\pm}$ in a clear way; hence, we can propose the 
phenomenological 
relations between them. As we will see in the 
application, for interfaces, general kinds of phenomenological 
relations need to be used, and for that reason we have 
written them like in Eq.~(\ref{F+}),
where the dependence of the functions $\vec{F}^{\pm}$, on the surface 
thermodynamic variables,
$T^{s}$, $n_{k}^{s}$, is implicit.

Equilibrium and non-equilibrium thermodynamic 
considerations impose some restrictions on the functions 
$\vec{F}^{\pm}$, 
namely:
\begin{itemize}
\item[a)] At equilibrium, all thermodynamic fluxes and forces must 
vanish, 
which
leads to the requirement that $\vec{F}^{\pm}$ have to vanish in 
equilibrium, 
that is, $\vec{F}^{\pm}( \vec{0}, \vec{0} ) = 0$.
\item[b)] They have to be defined in such a way as to make surface
entropy production 
 positively definite, in accordance with the second law of 
thermodynamics,
that is,
$\vec{X}^{+} \cdot \vec{F}^{+} + \vec{X}^{-} \cdot \vec{F}^{-} \ge 0$
(see Eqs.~(\ref{metentr1}) and~(\ref{F+}) ).
\item[c)] Near equilibrium, in which linear phenomenological relations
hold,
the matrix of phenomenological coefficients satisfy Onsager's 
reciprocal
 relations.
 This matrix is precisely the Jacobian matrix for the functions 
$\vec{F}^{\pm}$ evaluated at equilibrium and, consequently, it must 
have the 
corresponding symmetry properties.
\end{itemize}
These three requirements are, in fact, the conditions expressed 
through 
Eqs.~(\ref{equil})-~(\ref{Onsager}).
 
\subsection{Particular cases}

\subsubsection{Steady conditions}

Under steady conditions, the model presented in subsection III.A, and 
in 
particular the boundary conditions, simplify considerably,  and
they can be written in a more useful form for practical applications. 
Indeed, 
from 
Eq.~(\ref{mas1dk}) we conclude
 that no  discontinuity in the conducting currents occurs across the 
 interface, that is ,
 $\vec{J}^{+} = \vec{J}^{-}$ at the interface. Hence, after 
identifying
  the terms of the
 right hand sides in Eq.~(\ref{F+}) and solving for one of 
 the two forces,
 we can write the boundary conditions in the form
 \begin{eqnarray}     
  \vec{X}^{-} & = & \vec{G}^{+} (\vec{X}^{+})  \label{bc1}\\
  \vec{J}^{+} & = & -\frac{1}{T} \vec{H}^{+} (\vec{X}^{+}) 
  \label{bc}
 \end{eqnarray}
 where $\vec{G}^{+}$ is a function satisfying the relation  
 \begin{equation}
 \vec{F}^{+} ( \vec{X}^{+}, \vec{G}^{+} (\vec{X}^{+}) ) =
 \vec{F}^{-} ( \vec{X}^{+}, \vec{G}^{+} (\vec{X}^{+}) ) 
 \label{G+}
 \end{equation}
 and $\vec{H}^{+}$ is given by
 \begin{equation}
 \vec{H}^{+} (\vec{X}^{+}) =
 \vec{F}^{+} ( \vec{X}^{+}, \vec{G}^{+} (\vec{X}^{+}) ) \, .
 \end{equation}
 It can be seen that the
 general restrictions mentioned previously require that
 $\vec{G}^{+} ( \vec{0} ) = \vec{H}^{+} (\vec{0}) =  0$ and that 
 \begin{equation}
 (\vec{X}^{+} + 
 \vec{G}^{+}( \vec{X}^{+} ) ) \cdot \vec{H}^{+} ( \vec{X}^{+} ) \ge 0
 \label{cond}
 \end{equation}
 for all values of $\vec{X}^{+}$. 
  
 This last relationship is more naturally expressed by using
 the discontinuities in the bulk electrochemical potentials,
 $\vec{[\phi]}_{-}$, as the independent variables.
 In this case, by taking into account  that 
 $\vec{[\phi]}_{-} = \vec{X}^{+} + \vec{X}^{-}$, we may rewrite 
Eqs~(\ref{bc1})
 and~(\ref{bc}) as
\begin{eqnarray}      
& &\vec{J}^{+} = - \frac{1}{T} \vec{F}_{1} (\vec{[\phi]}_{-})  
\label{F1} \\
& &\vec{X}^{+}= \vec{F}_{2} (\vec{[\phi]}_{-})  \label{F2}
\end{eqnarray}
where $\vec{F}_{2}$ satisfies 
\begin{equation}
 \vec{G}^{+}( \vec{F}_{2} (\vec{[\phi]}_{-})) + 
\vec{F}_{2}(\vec{[\phi]}_{-}) = \vec{[\phi]}_{-} 
\label{G}
\end{equation}
 and  $\vec{F}_{1}$
 \begin{equation}
 \vec{F}_{1}(\vec{[\phi]}_{-}) = 
\vec{H}^{+} (\vec{F}_{2}(\vec{[\phi]}_{-})) \, .
\label{H}
\end{equation}
Now the thermodynamic restrictions are simply
\begin{eqnarray}
\vec{F}_{2}(\vec{0}) = \vec{F}_{1}(\vec{0}) = \vec{0} \label{res1} \\
\vec{[\phi]}_{-} \cdot \vec{F}_{1}(\vec{[\phi]}_{-}) \ge 0 \, . 
\label{res2}
\end{eqnarray}
 The steady state boundary conditions expressed through 
Eqs.~(\ref{F1}) 
and~(\ref{F2}), have the appropriate form to be used in practical 
calculations,
as we will show in future works.
\subsubsection{No surface states}

Our purpose in this subsection is to present the simplified version
of the model corresponding to the case in which the surface number
densities vanish, that is , when  $n_{i}^{s} = 0$ for $i=1,2,3$. This 
case
may be related to the case in which there are not surface, or 
interfacial,
states. The first simplification comes from the fact that now we 
obviously,
do not
need a surface state equation like Eq.~(\ref{stat1s}). 
Moreover, from the surface continuity equations, Eq.~(\ref{mas1dk}), 
we can
see that the number density currents are continuous across the 
interface,
that is, $[\vec{J}]_{-} = 0$. Finally, it can be shown that
 the entropy production corresponding
to this situation can be obtained from Eq.~(\ref{s1}) by simply 
imposing $[\vec{J}]_{-} = 0$ on it. One then arrives at,
\begin{equation}
\sigma_{s}^{s} = - \frac{1}{T} [\vec{\phi}]_{-} \cdot \vec{J}^{+} \, .
\label{nosur}
\end{equation}
From this expression we see that, in this case, we obtain a {\em 
single}
 discontinuity
relationship, that can 
be written as
\begin{equation}
\vec{J}^{+} = - \frac{1}{T} \vec{M} ( [\vec{\phi}]_{-} ) \label{disno}
\end{equation}
instead of the {\em two} obtained in the general case Eq~(\ref{F+})
or Eqs~(\ref{F1}) and~(\ref{F2})  in the steady case.        
As before, $\vec{M}$ is subject to some thermodynamic restrictions, 
namely
\begin{eqnarray}
& \mbox{} & \vec{M} (\vec{0})  =  0 \label{resno1} \\
& \mbox{} & [\vec{\phi}]_{-} \cdot \vec{M} ( [\vec{\phi}]_{-} ) \geq 0
                                                 \label{resno2} \\
& \mbox{} & \left. \frac{\partial M_{1}}{\partial [\phi_{2}]_{-}} 
\right|_{eq} = 
\left. \frac{\partial M_{2}}{\partial [\phi_{1}]_{-}} \right|_{eq} 
\end{eqnarray}
coming from the equilibrium condition, the second principle of
thermodynamics and the Onsager's symmetries. 

\section{Application of the model}

In this section we will first show, in a more or less
heuristic way, that the model presented
along this paper is complete (see below). For simplicity's
sake we will focus on unipolar systems, but similar arguments can be 
applied to bipolar systems. 
As an example of application, we will 
show that the thermionic 
emission-diffusion
theory for metal-semiconductor (M-S) abrupt junctions~\cite{Sze}, is a
thermodynamically consistent model, because it can be
fitted into the framework of the present model. Similar arguments
may be used to analyze more complicated situations,
in which not only thermodynamic consistency is expected to be proven, 
but
also a generalization of the corresponding models will be derived.
                                                                   
\subsection{Completeness of the model}

What will be proven in this subsection is that if the unspecified 
phenomenological
equations, Eqs.~(\ref{stat1s}), and~(\ref{F+}) in the general case or
Eq.~(\ref{disno}) in the non-surface state case, are given explicitly 
and the {\em physical conditions at infinity}
are specified, then we can find the solution of the model. By physical
conditions at {\em infinity}, we mean those conditions
that one assumes the solution must satisfy far away from the 
interface.
For instance, for infinite systems, one assumes that  the interfacial 
effects
die out at infinity, and hence that the system is locally neutral 
there,
that is, $n_{e}^{\pm} \rightarrow n_{0}^{\pm}$ and 
$\frac{\partial n_{e}^{\pm}}{\partial x}  \rightarrow 0$
for $ x \rightarrow \pm \infty$, where we have assumed the single 
carriers
to be electrons (subindex $e$) and the donor density (subindex 0) to 
be uniform.
For finite systems, one assumes that, at a certain distance away from 
the
interface, the interfacial effects have nearly died out, and, 
therefore,
one searches for an approximate solution that satisfies this 
condition.
                                                              
In any situation, our model provides the corresponding solutions. To 
see this,
let $x_{0}^{\pm}$ be,
for the moment, two unspecified points, at which the neutrality 
conditions
are satisfied within a certain degree of approximation, that is,
\begin{eqnarray}
                             n_{e}^{\pm}(x_{0}^{\pm}, t) & = &
n_{0}^{\pm} + \Delta_{1}^{\pm} \label{ap1} \\
\frac{\partial n_{e}^{\pm}(x_{0}^{\pm}, t) }{\partial x} & = & 
\Delta_{2}^{\pm}
 \label{ap2}
\end{eqnarray}
where $\Delta_{1,2}^{\pm}$ take small values ( for infinite
systems these precise values are not 
important, for at the end of the argument one tends them to zero).
Note that by taking Eqs.~(\ref{ap1}) and~(\ref{ap2}) as boundary 
conditions, one can solve the semiconductor bulk equations, 
Eqs.~(\ref{contk}),
 ~(\ref{qPo}), ~(\ref{driftd}). A simple way to do this is to
note that
that these equations
can be transformed into a single equation for the electric
field, see for instance \cite{Bonilla}, 
\begin{equation}
\frac{\partial E^{\pm}}{\partial t} 
  - D_{e}^{\pm} \frac{\partial^{2} E^{\pm}}{\partial x^{2}} 
  - \tilde{\mu}^{\pm} E^{\pm} \left( \frac{\partial E^{\pm}}{\partial 
x}
  - \frac{e}{\epsilon^{\pm}} n_{0}^{\pm} \right) = 
\frac{I}{\epsilon^{\pm}}
\label{camp}
\end{equation}
where $I$ is the total electric density current, and that 
Eqs.~(\ref{ap1}) and~(\ref{ap2}) can be rewritten as,
\begin{eqnarray}
\frac{\partial E^{\pm}(x_{0}^{\pm}, t)}{\partial x} & = & 
\tilde{\Delta}_{1} \label{Bc1} \\
                            E^{\pm}(x_{0}^{\pm}, t) & = &
   \frac{1}{\sigma^{\pm}} J_{e}^{\pm}(x_{0}^{\pm}, t) + 
\tilde{\Delta}_{2}                                      
\label{bc2}
\end{eqnarray}
where Eqs.~(\ref{qPo}) and~(\ref{driftd}) have been used. In 
Eqs.~(\ref{Bc1})
and~(\ref{bc2}), $\sigma^{\pm} =  e \tilde{\mu}^{\pm} n_{0}^{\pm}$ are
the bulk 
conductivities, and $\tilde{\Delta}_{1,2}^{\pm}$, two very small
quantities, related to $\Delta_{1,2}^{\pm}$.

Once one has solved for the semiconductor bulk equations,
(usually numerically),
one can compute some useful quantities like: $J_{e}^{\pm} (0,t)$, 
$n_{e}^{\pm} (0,t)$,
$\mu_{e}^{\pm} (0,t)$, $V^{\pm} (0,t) - V^{\pm} (x_{0}^{\pm},t)$,
$\phi_{e}^{\pm} (0,t) - \phi_{e}^{\pm} (x_{0}^{\pm},t)$, ... by simply
using 
 the bulk equations again. Note that all these quantities will 
obviously be 
 dependent on $x_{0}^{\pm}$, $I(t)$ and $\tilde{\Delta}_{1,2}^{\pm}$.
If surface states are present, one can then compute the corresponding
expressions 
for the surface quantities, $q^{s}(t)$ and $\mu_{e}^{s}(t)$
as functions of
$x_{0}^{\pm}$, $I$ and $\tilde{\Delta}_{1,2}^{\pm}$ . This can be done
as follows: from Poisson's equation,
Eq.~(\ref{qPo}), and the discontinuity equation, Eq.~(\ref{discP}), 
one has
\begin{equation}
E^{+} (x_{0}^{+}, t) - E^{-} (x_{0}^{-}, t) = q^{-} (t) + q^{+} (t) + 
q^{s} (t)
\label{ara}
\end{equation}
where we have defined
\begin{equation}
q^{\pm} = \pm \frac{e}{\epsilon^{\pm}} \int_{0}^{x_{0}^{\pm}} 
                        (n_{0}^{\pm} - n_{e}^{\pm}) d x  \, .
\end{equation}
From Eq.~(\ref{ara}) one obtains $q^{s}(t)$, for it is the only 
unknown 
quantity
in this expression. By using the surface state equation, 
Eq.~(\ref{stat1s}), 
one can
then derive the corresponding expression for $\mu_{e}^{s}(t)$ 
(we assume that the surface
donor density, $n_{0}^{s}$, as well as the bulk ones, $n_{0}^{\pm}$, 
is known).

 All the previous results suffice to derive
 the expressions for $X_{e}^{\pm}(0,t)$ and
 $J_{e}^{\pm}(0,t)$  in terms of $x_{0}^{\pm}$, $I$ and 
 $\tilde{\Delta}_{1,2}^{\pm}$. By substituting them into
  the discontinuity
 relationships, Eq.~(\ref{F+}), one then obtains a system of two 
equations,
 which allows one to derive the expressions for $x_{0}^{\pm}$ as 
functions 
of $I$ and $\tilde{\Delta}_{1,2}^{\pm}$.
At this point one is then able to eliminate
the dependence of all the previous
results on $x_{0}^{\pm}$. As a result, all these magnitudes become 
only 
functions of $I$ and $\tilde{\Delta}_{1,2}^{\pm}$.
 
For an infinite system, one then tends $\tilde{\Delta}_{1,2}^{\pm}$ to
zero, 
(and consequently $x_{0}^{\pm}$ will tend to $\pm \infty$). The 
resulting
   expressions, which are now only functions of $I$, constitute
the
          searched solution of the model.
          
For a finite system, one assumes that the results found with finite 
values for
 $\tilde{\Delta}_{1,2}^{\pm}$, 
are the (approximate) solutions for the active region of the device, 
and
that the remaining parts of the system behave as in the absence of 
the interface. We then arrive at the desired solution in this case.
In some cases, for instance for short devices,
this is not a good approach, and a more elaborated one must be carried
out~\cite{Henisch}.
In any case, it can be shown that the correct implementation of the 
model
gives the corresponding solution.

For the particular case in which there are no surface states
present, the same derivation holds, by simply noting that
in this case one has $q^{s} (t) = 0$. Therefore,
 Eq.~(\ref{ara}) constitutes the first equation to determine 
$x_{0}^{\pm}$, and the second one is the single discontinuity 
relation for that case, Eq.~(\ref{disno}).

\subsection{Thermionic emission-diffusion theory for M-S junctions}

The thermionic emission-diffusion theory for 
abrupt M-S junctions without interface states is a currently
accepted model to describe the rectifying properties
of M-S junctions when the main limiting
current mechanisms are the thermionic emission of carriers
over the junction barrier and the diffusion of the carriers inside 
the semiconductor~\cite{Sze},~\cite{Roderic}. Our purpose in this
subsection is to show that
this model can  fit into the framework of the model presented 
throughout this paper,
and that, as a consequence, it is a thermodynamically consistent 
model.

As in many cases, this model assumes drift-diffusion 
transport equations for the bulks~\cite{Sze},  which are
equivalent to the
ones presented in section III. On the other hand, it also
assumes the following boundary condition to describe the
thermionic emission processes,~\cite{Sze1} (see also~\cite{Sze})
\begin{equation}
e J_{e}^{+} = V_{R} (n_{m} - n_{e}^{+})
\label{bcS}
\end{equation}
where $V_{R}$ is a positive constant, and $n_{m} = 
N_{C}^{+} e^{- \frac{e}{k_{B} T} \phi_{b}^{+}}$. This boundary 
condition
is a thermodynamically consistent, in the sense of the present paper. 
Indeed,
 by using Eq.~(\ref{nonde}) it can be rewritten 
in the form of a discontinuity relation, 
\begin{equation}
 -e J_{e}^{+} = I_{T} e^{-\frac{e}{k_{B} T} \phi_{b}^{+}}
 \left( e^{\frac{1}{k_{B} T} [\phi_{e}]_{-}} - 1 \right)
 \label{tebc}
\end{equation}
where $I_{T} = V_{R} N_{C}^{+}$. Furthermore,
 this discontinuity relation clearly
 satisfies the thermodynamic restrictions derived for
that case,
Eqs.~(\ref{resno1}) and~(\ref{resno2}) (for 
unipolar systems without interface states there are no symmetry 
relationships). 
 It is worth noting
that, although Eq.~(\ref{resno1}) is expected to be satisfied for
all the models, Eq.~(\ref{resno2}) might not, for its origin is purely
thermodynamic, and it has not been included, as far as we know, in any
phenomenological model.
Moreover, it is 
should be pointed out that the constitutive relation given through 
Eq.~(\ref{tebc}) is
effectively non-linear (as we have assumed to be), in the sense that 
it 
does not relate the
current and the force linearly. However, this does not constitute
a singular case in non-equilibrium thermodynamics for there are
different irreversible processes
(chemical reactions~\cite{Degro}, adsorption 
processes~\cite{Ignasi},...)
 which display such non-linear behaviours.

Similar analysis may be carried out for more complicated models, for 
instance
those including interface states. By fitting those models into the 
framework
of our model their thermodynamic consistency may be established and,
what is more important, their generalization eventually derived. These
facts
illustrate the power of the application of the thermodynamic methods 
 to semiconductor systems.
                                                 
\section{Conclusions}

In this paper we have presented a fully macroscopic, and 
thermodynamically
 consistent, derivation
 of a phenomenological one-dimensional model for drift-diffusion
 processes in a junction. The model, which consists of the usual 
transport
 equations for the bulk systems that conform the junction, 
 also incorporates the appropriate boundary conditions or 
discontinuity
 relationships, and the
 equations describing the surface phenomena. These last two 
ingredients
 are essential from a fundamental point of view, and they allow
 a complete modelling of semiconductor devices, in which junctions
 play a so relevant role.
 
 The derivation of the model has been completely based on 
thermodynamic
 arguments. A first step has consisted of thermodynamically 
characterising the
 junction, by means of the introduction of the surface and 
(extrapolated) bulk
 variables. Therefore, the methods of non-equilibrium thermodynamics 
for 
 systems
 with interfaces~\cite{Bedo2}, have
 been applied giving rise directly to the above-mentioned model. It is
worth 
 noting
 that this derivation clearly ensures the thermodynamic consistency
 of the whole model.
 Of special interest are the thermodynamic restrictions
 affecting the drift-diffusion boundary conditions, 
 Eqs.~(\ref{equil})-~(\ref{Onsager}),
  for they must be satisfied for any proposed boundary condition.
  
 As in any phenomenological model, explicit expressions for some 
functions or
 parameters are not given by the theory itself. Only when the model is
 applied to several different systems, or when the appropriate 
microscopic 
 models are proposed, information about these unspecified parts may be 
obtained.
 It should be pointed out, that, in any case, these results must 
satisfy 
 the general restrictions imposed by the thermodynamic derivation.
 
  We have shown, in the particular case of unipolar systems,
 that the model is complete, in the sense that, once the physical
 conditions at {\em infinity} are specified, it completely
 characterizes the system. A similar derivation can be developed for
 bipolar systems.
 
 As a simple application, we have analysed the thermionic 
emission-diffusion
 theory for one-dimensional abrupt M-S junctions without interface
 states and have shown
 that this theory is thermodynamically consistent, in the sense of the
 present paper,
 by showing that it fits well into the framework of the present model.
 In addition, we have also justified with this example the use of 
non-linear 
 constitutive
 relations in the formulation of the discontinuity relations.
 Similar procedures can be applied to more complicated situations, 
allowing
 one to extract very useful information on the semiconductor systems.

 Finally, it should be noted that the drift-diffusion model presented 
here
 may be 
 generalised  in a more o less systematic
 way, by including other irreversible processes or by considering 
 three-dimensional
 systems, which would allow us to study, for instance,
  inhomogeneous junctions
 ~\cite{Tung},~\cite{Sullivan}. Among these generalisations, it is 
 especially important to include 
  recombination-generation processes or tunnelling processes in the 
model
 because their relevance to semiconductor devices is well known. 
 The treatment of recombination-generation processes in the framework
 of non-equilibrium thermodynamics
 will be considered in 
 future papers. On the other hand, the non-local nature of the 
tunnelling
 processes
  makes their implementation in this (local) framework
  rather complicated, although some 
 local macroscopic 
 formulations  are present in the current 
 literature~\cite{Ancona2}, ~\cite{Siam}. 

\acknowledgements

It is a pleasure to acknowledge Prof. L.L. Bonilla and I. 
Pagonabarraga
for fruitful discussions.
This work has been supported by DGICYT of the Spanish Government"
under grant PB92-0859 and the European Union Human Capital and
Mobility Programme, contract ERB-CHR-XCT93-0413. One of us (G.G.) 
wishes
to thank CIRIT of Generalitat de Catalunya for financial support.
 
\appendix
\section{On non-equilibrium thermodynamics of systems with interfaces}
\setcounter{equation}{0}
 Our purpose in this appendix is to establish the non-equilibrium 
 thermodynamics basis of multicomponent electrically conducting 
polarizable 
 systems divided by an interface by following the procedure indicated 
 in~\cite{Bedo}.
 Crucial in this analysis is the derivation of the surface and bulk 
entropy 
 productions
 accounting for the irreversible processes occurring in the system. 
The
 phenomenological equations derived from these quantities correspond 
to the 
 constitutive 
 equations for the interfacial and bulk currents and to the boundary 
conditions.

 First of all, we will indicate the way to derive balance
 equations at  interfaces~\cite{Bedo2}. To this purpose, let us focus 
on 
 an unspecified
 quantity $d$ defined in the whole system. Its evolution is dictated 
by the 
 balance equation
  \begin{equation} 
 \frac{\partial}{\partial t} d_{V} + div \vec{J}_{d}  =
                                                     \sigma_{d} 
  \label{bald+}
   \end{equation}
 where $\vec{J}_{d}$ and $\sigma_{d}$ are the corresponding current 
and 
 production, and where the subindex $V$ means that this density is per
 unit of volume.
 Now, using  the decomposition of the fields in terms of the bulk 
fields and 
 excess fields given through Eq.~(\ref{dexces}) we arrive at
 \begin{eqnarray}
& &\left(\frac{\partial}{\partial t} d_{V}^{+}    + div 
\vec{J}_{d}^{+}  -
                           \sigma_{d}^{+} \right) \theta^{+} +  
\nonumber \\
& &\left(\frac{\partial}{\partial t} d_{V}^{-}    + div 
\vec{J}_{d}^{-}  -
                           \sigma_{d}^{-} \right) \theta^{-} +   
\nonumber \\
& &\left(\frac{\partial}{\partial t} d_{ex}    + div \vec{J}_{d,ex}  
                         + \hat{n} \cdot   [ \vec{J}_{d} ]_{-} \, 
\delta^{s} -
                                         \sigma_{d,ex} \right) = 0
 \label{delta}                                                   
\end{eqnarray}                                                      
where $\delta^{s}$ is Dirac's delta function coming from the 
derivatives 
of Heaviside's functions~\cite{Bedo2}, $\hat{n}$ is the unit normal 
vector
pointing from $-$ to $+$ and $[A]_{-} = 
A^{+}(0,y,z)-A^{-}(0,y,z)$ (we are assuming that the interface is at 
rest 
and is identified with the plane  $x=0$).
In the bulk, $d^{\pm}$ obeys the balance equations
\begin{equation}
\frac{\partial}{\partial t} d_{V}^{\pm} + div \vec{J}_{d}^{\pm}  =
                                           \sigma_{d}^{\pm}
\end{equation}
As a consequence, the two first lines in Eq.~(\ref{delta}) vanish. By
averaging the remaining third term
as well as using the definition of the surface fields given in 
Eq.~(\ref{surface}) we 
arrive at the surface balance equation
\begin{equation} 
\frac{\partial}{\partial t} d_{A}^{s}    + div 
\vec{J}_{d,\parallel}^{s}    
          + \hat{n} \cdot   [ \vec{J}_{d} ]_{-} = \sigma_{d}^{s}    
\label{balds}
\end{equation}
where the subindex $A$ refers to the fact that the density is per unit
of area, and
where only the parallel component of the current appears because the
integration
of the normal component vanishes. Usually one defines a surface flux 
with no
normal component and calls this restriction a transversality 
condition.
Further details may be seen in~\cite{Bedo2}.

We will now apply the former scheme in order to obtain 
the set of balance equations 
describing the evolution of the relevant quantities in the system.
For the balance of mass of the $k$-th component one has
\begin{equation} 
\frac{\partial}{\partial t} \rho_{k}^{\pm} + div \vec{J}_{k}^{\pm} =
                                                  \sigma_{k}^{\pm} 
\hspace{0.5 cm} \mbox{;} \hspace{0.5 cm} k = 1, 2, 3   
\label{balm+}
\end{equation}
                                                    
\begin{equation} 
\frac{\partial}{\partial t} \rho_{k}^{s} + div 
\vec{J}_{k,\parallel}^{s} 
                 + \hat{n} \cdot   [ \vec{ J}_{k} ]_{-} = 
\sigma_{k}^{s}
 \hspace{0.5 cm} \mbox{;} \hspace{0.5 cm} k = 1, 2, 3                 
\label{balms}
 \end{equation}
where $\vec{J}_{k}^{\pm}$ and $\vec{J}_{k}^{s}$ are the diffusion 
currents and 
$\sigma_{k}^{\pm}$ and $\sigma_{k}^{s}$ the production rates. From 
these
equations one may derive the balance equations for the total mass 
$\rho^{\pm} = \sum_{k=1}^{3} \rho_{k}^{\pm}$,
$\rho^{s} = \sum_{k=1}^{3} \rho_{k}^{s}$
 and total charge, $q_{V}^{\pm} = \sum_{k=1}^{3} z_{k} 
\rho_{k}^{\pm}$,
$q_{A}^{s} = \sum_{k=1}^{3} z_{k} \rho_{k}^{s}$, with $z_{k}$ being 
the
charge-mass ratio of the k-th component,
\begin{equation} 
\frac{\partial}{\partial t} \rho^{\pm} + div \vec{J}_{\rho}^{\pm} = 0
\label{baltm+}
\end{equation}
                                                    
\begin{equation} 
\frac{\partial}{\partial t} \rho^{s} + div 
\vec{J}_{\rho,\parallel}^{s} 
                 + \hat{n} \cdot   [ \vec{ J}_{\rho} ]_{-} = 0
\label{baltms}
\end{equation}
\begin{equation} 
\frac{\partial}{\partial t} q_{V}^{\pm} + div \vec{J}_{c}^{\pm} = 0  
\label{baltq+}
\end{equation}
                                                    
\begin{equation} 
\frac{\partial}{\partial t} q_{A}^{s} + div \vec{J}_{c,\parallel}^{s} 
                 + \hat{n} \cdot   [ \vec{J} ]_{-} = 0
\label{baltqs}
\end{equation}
where we have taken into account the reaction conservation relations 
$\sum_{k=1}^{3} \sigma_{k}^{\pm} = 0$, $\sum_{k=1}^{3} z_{k} 
\sigma_{k}^{\pm} = 0$, $\sum_{k=1}^{3} \sigma_{k}^{s} = 0$, 
$\sum_{k=1}^{3} z_{k} \sigma_{k}^{s} = 0$, and we have defined the 
total
mass currents as $\vec{J}_{\rho}^{\pm} = \sum_{k=1}^{3} 
\vec{J}_{k}^{\pm}$,
$\vec{J}_{\rho, \parallel}^{s} = \sum_{k=1}^{3} \vec{J}_{k, 
\parallel}^{s}$
and the charge conducting currents as 
$\vec{J}_{c}^{\pm} = \sum_{k=1}^{3} z_{k} \vec{J}_{k}^{\pm}$,
$\vec{J}_{c, \parallel}^{s} = \sum_{k=1}^{3} z_{k} 
\vec{J}_{k,\parallel}^{s}$.

In the same way for the balance of momentum one has
\begin{equation} 
\frac{\partial}{\partial t} \vec{g}_{V}^{\pm}    + div P^{\pm} =
                                    \vec{f_{V}}^{\pm}   
\label{balg+}
 \end{equation}
\begin{equation} 
\frac{\partial}{\partial t} \vec{g}_{A}^{s} 
+ div P_{\parallel}^{s} + \hat{n} \cdot [ P ]_{-}  = 
\vec{f_{A}}^{s}
\label{balgs}
 \end{equation}
where $\vec{g}_{V}^{\pm}$ and $\vec{g}_{A}^{s}$ are the momentum 
densities,
 $P^{\pm}$ and $P^{s}$, the pressure tensors 
 and $\vec{f}_{V}^{\pm}$ and 
$\vec{f}_{A}^{s}$ the density force fields. Precise expressions for 
these 
quantities
can be found in 
     ~\cite{Degro} for the bulk and in      ~\cite{Bedo} for the 
interface.

Finally, the balance for the total energy is given by
\begin{equation} 
\frac{\partial}{\partial t} e_{V}^{\pm} + div \vec{J}_{e}^{\pm} = 0 
\label{bale+}
 \end{equation}
                                                     
\begin{equation} 
\frac{\partial}{\partial t} e_{A}^{s}  + div \vec{J}_{e,\parallel}^{s}
                         + \hat{n} \cdot   [ \vec{J}_{e} ]_{-}= 0  
  \label{bales}
 \end{equation}
where $\vec{J}_{e}^{\pm}$ and $\vec{J}_{e}^{s}$ are the energy 
currents. The  internal
energy equations follow from the previous equations 
provided the total 
energy is expressed in terms of the internal energy and kinetic and 
potential
contributions. One then arrives at
\begin{equation} 
\frac{\partial}{\partial t} u_{V}^{\pm} + div \vec{J}_{u}^{\pm}    =
                                     \sigma_{u}^{\pm}   
\label{balu+}
 \end{equation}
                                                     
\begin{equation} 
\frac{\partial}{\partial t} u_{A}^{s}  + div \vec{J}_{u,\parallel}^{s}
               + \hat{n} \cdot   [ \vec{J}_{u} ]_{-}=
                                    \sigma_{u}^{s}    
  \label{balus}
 \end{equation}
where $\vec{J}_{u}^{\pm}$ and $\vec{J}_{u}^{s}$ are the internal 
energy current 
and $\sigma_{u}^{\pm}$ and $\sigma_{u}^{s}$
are the internal energy productions (see ~\cite{Degro} and~\cite{Bedo}
for 
explicit expressions of these quantities). 

In order for the system of balance equations to become a complete set 
of
 differential 
equations for the fields, we need explicit expressions for the 
currents 
appearing
in them. These can be obtained in the framework of non-equilibrium 
thermodynamics.

In fact, after assuming local equilibrium, we then propose the 
fundamental
equations in the bulks and at the surface,
\begin{equation}
  s_{V}^{\pm} = f^{\pm} ( u_{V}^{\pm} , \rho_{k}^{\pm} ,
                          \vec{P}_{V}^{\pm},\vec{M}_{V}^{\pm} ) 
   \label{Gibbs+}
   \end{equation}
\begin{equation} 
 s_{A}^{s}  = f^{s} ( u_{A}^{s}  ,\rho_{k}^{s} ,
                          \vec{P}_{A}^{s} ,\vec{M}_{A}^{s} ) 
  \label{Gibbss}
  \end{equation}
 (or Gibbs equations when they are expressed in differential form)
 to hold locally.
After derivation with respect to time, one then obtains the 
following equations in the bulks
\begin{eqnarray}
 \frac{\partial}{\partial t} s_{V}^{\pm} & = &
        \frac{1}{T^{\pm}} \frac{\partial}{\partial t} u_{V}^{\pm}
       -\sum_{k=1}^{3} \frac{\mu_{k}^{\pm}}{T^{\pm}} 
                                \frac{\partial}{\partial t} 
\rho_{k}^{\pm}  \nonumber \\
                                         &   & -
       \frac{\vec{E}_{eq}^{\pm '}}{T^{\pm}} \cdot   
                           \frac{\partial}{\partial t} 
\vec{P}_{V}^{\pm '}
       -\frac{\vec{B}_{eq}^{\pm '}}{T^{\pm}} \cdot   
                           \frac{\partial}{\partial t} 
\vec{M}_{V}^{\pm '}  
\label{Gibbs1+}
\end{eqnarray}
and at the interface
 \begin{eqnarray} 
 \frac{\partial}{\partial t} s_{A}^{s} & = &
       \frac{1}{T^{s}} \frac{\partial}{\partial t} u_{A}^{s}
      -\sum_{k=1}^{3} \frac{\mu_{k}^{s}}{T^{s}}
                                \frac{\partial}{\partial t} 
\rho_{k}^{s} \nonumber \\
                                       &   & -
       \frac{\vec{A}_{1 eq}^{s '}}{T^{s}} \cdot   
                              \frac{\partial}{\partial t} 
\vec{P}_{A}^{s '}    
      -\frac{\vec{A}_{2 eq}^{s '}}{T^{s}} \cdot  
                              \frac{\partial}{\partial t} 
\vec{M}_{A}^{s '}
\label{Gibbs1s}
 \end{eqnarray}
In~(\ref{Gibbs1+}) the primes refer to the bulk baricentric 
 velocity and in~(\ref{Gibbs1s}) to the surface baricentric one. 
Moreover
 the thermodynamic definitions of the intensive parameters have been 
used.
 
 Following~\cite{Bedo} we will take 
 \begin{eqnarray}
 \vec{A}_{1}^{s} = ( [D_{n}]_{+} , [\vec{E}_{\parallel}]_{+}) \\
 \vec{A}_{2}^{s} = ( [B_{n}]_{+} , [\vec{H}_{\parallel}]_{+})
 \end{eqnarray}
 as the conjugate variables of the surface polarisations.
 Then, by inserting the expressions
  for the balance equations
 previously stated in Eqs.~(\ref{Gibbs1+}) and~(\ref{Gibbs1s}), one 
arrives at
  the expressions for the bulk and surface
 entropy balance equations
 \begin{equation} 
 \frac{\partial}{\partial t} s_{V}^{\pm}    + div \vec{J}_{s}^{\pm}   
  = \sigma_{s}^{\pm} 
  \label{bals+}
   \end{equation}
                                                     
\begin{equation} 
\frac{\partial}{\partial t} s_{A}^{s}    + div 
\vec{J}_{s,\parallel}^{s}    
                         + \hat{n} \cdot   [ \vec{J}_{s} ]_{-}=
                                                     \sigma_{s}^{s} 
 \label{balss}
 \end{equation}
where $\vec{J}_{s}^{\pm}$ and $\vec{J}_{s}^{s}$ are the entropy 
currents
and $\sigma_{s}^{\pm}$ and $\sigma_{s}^{\pm}$ are the bulk and 
surface  entropy
productions. The explicit calculations for $\vec{J}_{s}^{\pm}$ and 
$\vec{J}_{s}^{s}$, and $\sigma_{s}^{\pm}$ and $\sigma_{s}^{s}$
 may be found in      ~\cite{Degro} and~\cite{Bedo}. For the 
drift-diffusion
model we are considering, which in particular neglects relaxation 
polarisation
phenomena, generation-recombination processes, thermal, magnetic and
viscousus effects, one then obtains~\cite{Degro}
\begin{equation}
\sigma_{s}^{\pm} =     
 -\frac{1}{T} \sum_{k=1}^{3} \vec{J}_{k}^{\pm '} \cdot  
 (grad \, \mu_{k}^{\pm} - z_{k} \vec{E}^{\pm})
\label{sigmaf+drift} 
\end{equation}
and 
\begin{equation}
\vec{J}_{s}^{\pm '} = \frac{1}{T^{\pm}} \left( \vec{J}_{u}^{\pm '} -
          \sum_{k=1}^{3} \mu_{k}^{\pm} \vec{J}_{k}^{\pm '} \right)
\end{equation}
where the primed fluxes are defined as $\vec{J}_{d}^{\pm '} = 
\vec{J}_{d}^{\pm}
- \vec{v}^{\pm} d^{\pm}$. Here $\vec{v}^{\pm}$ is the baricentric 
velocity
defined through 
$\rho^{\pm} \vec{v}^{\pm} = \vec{J}_{\rho}^{\pm}$. For a
one-dimensional system, Eq.~(\ref{sigmaf+drift}) is precisely 
Eq.~(\ref{elect+})
used in Section III.B. 

A similar, but much more complicated, analysis
 provides the expressions for the surface flux and entropy 
production~\cite{Bedo}. For a one-dimensional system no equation for 
the flux
is necessary. For the entropy production, in the drift-diffusion 
approximation,
one obtains ~\cite{Bedo},
\begin{equation}
\sigma_{s}^{s} = -\frac{1}{T^{s}} \sum_{k=1}^{2} [J_{k} (\mu_{k} - 
           \mu_{k}^{s})]_{-} + 
           \frac{1}{T} E_{n}^{s} [\sum_{k=1}^{2} z_{k} J_{k}]_{+}
\label{s1A}           
\end{equation}
where terms proportional to the third power of velocity have been 
neglected,
and where $E_{n}^{s}$ is the surface variable corresponding to the 
normal component of the electric field. In this paper we have assumed
that the surface is not polarizable, that is $P_{A,n}^{s} = 0$. Due 
to the
relation, see~\cite{Bedo}, $P_{A,n}^{s} = - E_{n}^{s}$, one then has 
 $E_{n}^{s} = 0$. In this case, Eq.~(\ref{s1A}) transforms into
Eq.~(\ref{s1}) used in Section III.B.
In some cases it is useful to take into account the surface 
polarisation,
\cite{Bedo5}. In these cases, one can show that the discontinuity in 
the 
electric potential is given by $[V]_{-} = -E_{n}^{s}$, and that one
needs an additional surface state equation, that can be written, for
instance as $P_{A,n}^{s} = P_{A,n}^{s} ([D]_{+})$. It is worth noting
that even in this case Eq.~(\ref{s1A}) can be rewritten as 
Eq.~(\ref{metentr1}).

\end{multicols}

\end{document}